\documentclass[amsmath,amssymb]{revtex4}
\usepackage{graphicx}
\usepackage{hyperref}
\usepackage{subfigure}
\pdfoutput=1
\begin{document}

\title{2-photon ionization and necessary laser and vacuum systems for experiments with trapped strontium ions}
\author{E. Kirilov and S. Putterman}
\affiliation{UCLA Department of Physics and Astronomy, Los Angeles,
California 90095, USA}

\begin{abstract}
We describe a efficient way to photoionize strontium atoms in a
linear radio-frequency trap. We use a 2-photon second order process
to excite the autoionization resonance
$(4d^{2}+5p^{2})^{\phantom{s}1}D_{2}$. A doubled pulsed Ti:Saphire
laser system is used at 431nm to provide 100fsec pulses at 82Mhz.
The fabrication of the laser systems for addressing the $Sr^{+}$
transitions necessary for laser cooling and excitation of quantum
jumps, vacuum system and ion trap structure are also described in
detail. With the current setup a easy and repeatable trapping of
linear ion chains is achieved at low $\thicksim8\times10^{-11}$Torr
pressures.
\end{abstract}
\pacs{78.60.Mq, 25.45.-z, 28.20.-v, 28.52.-s}
 \maketitle
  \textbf{1.
Introduction}

 The ability to trap individual isolated ions provides an opportunity to experimentally test
 the foundations of quantum mechanics. Observation of quantum jumps~\cite{Dehmelt86,Wineland86}
 for a single ion using the Dehmelt shelving scheme~\cite{Dehmelt75}
  yield a means of testing, with a single resettable degree of freedom, on the inherent randomness of
  quantum mechanics~\cite{Erber85,Putterman85,berkeland04}.
An investigation of the consequences of a purely quantum
 phenomena as quantum teleportation~\cite{Bennett93,Kimble98},
 entanglement~\cite{wineland98}, especially as they relate to the
 goals of quantum computing has been facilitated by geometries where a larger number of ions can be trapped and stabilized.

In this paper a detailed description of the construction of the,
vacuum system, imaging system, feedback loops, trap designs and
laser systems needed to Doppler cool and trap a single strontium ion
~\cite{berkeland2002} are presented. In order to obtain greater
control over the seeding of ions into the trap, 2 photon resonance
enhanced photoionization is employed. For the technique presented
both photons have identical energy (431nm) and the intermediate
level is a virtual state. Both quadrupole~\cite{Wuerker59} and
linear traps are discussed. Linear traps have the advantage of
minimizing micromotion when multiple ions are
present~\cite{Raizen92,berkeland2002}.

  Two particle entangled states have been realized in both cavity
  QED~\cite{Haroche97} and in ion traps~\cite{Turchette98}. In cavity QED information is transferred between the electronic states of various
  atoms via the modes of the cavity. Challenges presented to this approach include, the short confinement times and the need to achieve high Q
  (so that the decoherence time is sufficiently long until the conditional dynamics is
  performed)~\cite{Kimble92}. In a ion trap containing multiple ions,
  the motional quantum levels are determined by the potential well
  as modified by the mutual coulomb interaction. These ladder states
  then serve as a memory for quantum information. The
  difficulty then is the
  achievement of a joint motional ground state and the problem with heating of the ions.

    A goal of this research effort is to combine the control facilitated by ion traps with the virtual excitation of
   a high-Q cavity~\cite{Guo00} which is then used as bus for quantum memory, as already
   proposed
   and pursued by several groups~\cite{Walter01,Blatt03,Blatt04}.

\textbf{2. Photoionization}
\begin{figure}[h]
\centering
\includegraphics[width=83mm]{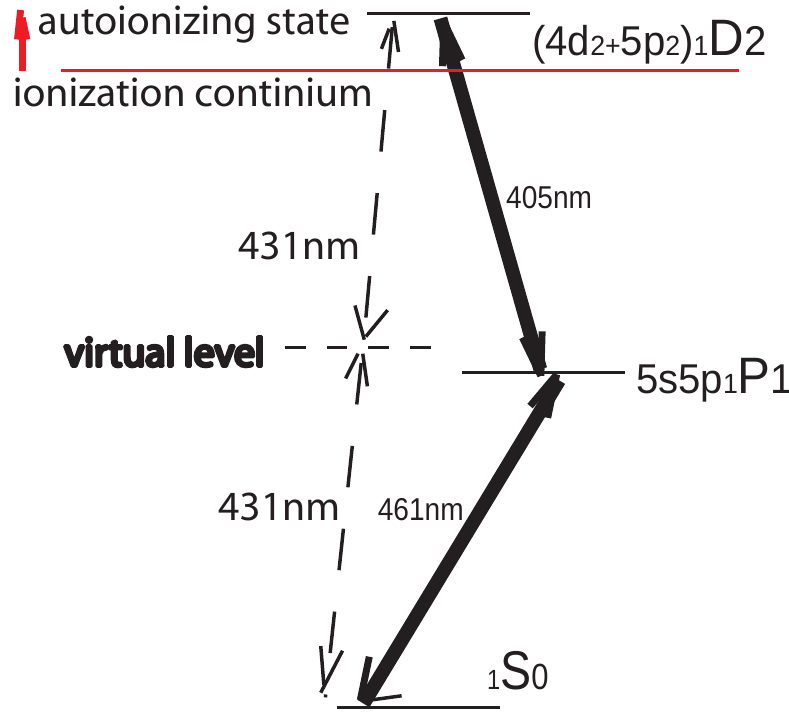}
\caption{Relevant level diagram of neutral Strontium used for the
photoionization method. In this work the 431nm light is supplied by
a 100fs, 0.5W average power pulses of a doubled TiSapphire laser. }
\end{figure}

 In these experiments Strontium ions are efficiently created and fed into the trapping region via two-photon
  resonance-enhanced photoionization~\cite{Baig98}. Compared to ionization via electron
bombardment two photon resonance enhanced photoionization has
various advantages. It's high efficiency means greatly reduced
atomic flux, less electrode contamination and the elimination of
problems due to charge accumulation on insulating surfaces, such as
the distortion of trap potentials~\cite{Drewsen00}. In addition the
vacuum remains nearly unchanged during the process of loading the
trap.

\begin{figure}[h]
\centering
\includegraphics[width=83mm]{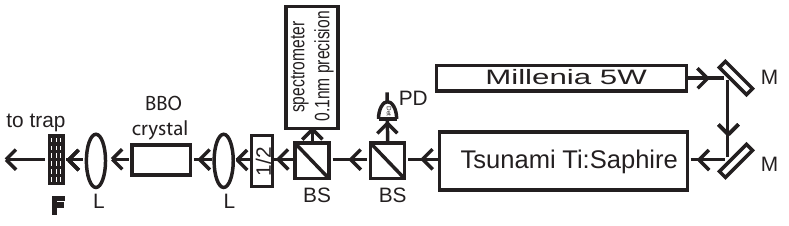}
\caption{Photoionizing system at 431nm.}
\end{figure}
   To ionize Sr a 431nm laser excites the transition
   $5s^{2\phantom{s}0}S_{0}\to(4d^{2}+5p^{2})^{\phantom{s}1}D_{2}$,
   where the upper level is an
   autoionizing resonance
   that is ~$470cm^{-1}$
   above the first ionization threshold (fig.1) ~\cite{Baig98}. The
   autoionizing state can also be reached
   with a two-step excitation using the intermediate $5s5p^{\phantom{s}1}P_{1}$
    level. This approach requires two lasers at 461nm and 405nm and the absolute cross-section for transition from
    $5s5p^{\phantom{s}1}P_{1}$ to
    $(4d^{2}+5p^{2})^{\phantom{s}1}D_{2}$ is 5.6$\times10^{-15}cm^{2}$~\cite{Mende95}.
    In our scheme the ionization cross-section is
    ~$10^{-26}cm^{4}/W$ at the peak of the autoresonance. This is a
    theoretical estimate based on an R-matrix approach combined with
    the multichannel quantum defect theory (MQDT)~\cite{Koenig98}. The two-photon
    transition amplitude from the initial state $\psi_{0}(E_{0})$ to
    the final $\psi_{f}(E_{f}=E_{0}+2\omega)$, using the lowest
    order in the perturbation theory is:
     \[
     T_{fo}^{(2)}=\lim_{\eta\to0+}\int\langle\psi_{f}|\textbf{D.e}|\varepsilon\rangle(E_{0}+\omega-\varepsilon+i\eta)^{-1}
     \langle\varepsilon|\textbf{D.e}|\psi_{0}\rangle
     \]
     where \textbf{e} is the electric field polarization and \textbf{D} is the
     electric-dipole operator.
\begin{figure}[h]
\centering
\includegraphics[width=123mm]{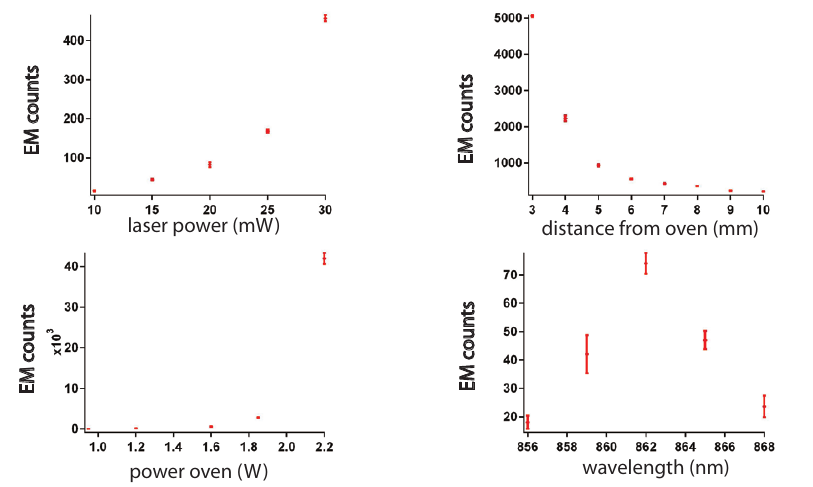}
\caption{Electron multiplier counts per second (used here instead of
arbitrary units) as a function of laser detuning, atomic oven power,
laser power and distance between the laser focal spot and atomic
oven aperture.}
\end{figure}
      The explicit summation over the discrete or continuum states is avoided using the Dalgarno-Lewis
      procedure~\cite{Lewis55}.
      The total cross section (in $cm^{4}W^{-1}$) is :
      \[
      \sigma'=\sigma/I=5.7466\times10^{-35}\sum_{j}|T_{fo}^{(2)}|^{2}
      \]
      where I is the laser intensity (in $Wcm^{-2}$) and the sum is
      performed over the possible total angular momentum states
      allowed by the Wigner-Echart theorem. In our case the laser is
      linearly polarized so both the $J=0$ and $J=2$ final states can be
      reached.

       The setup is shown on figure 2. A Tsunami laser (Spectra-Physics) is used to provide 100fsec pulses at
       82Mhz.
       The average power is $\thicksim$1/2W and the spectral width is
       $\thicksim$6nm. The output at 862nm is doubled with a BBO (2mm) crystal
       in a single pass giving ~50mW of average power at 431nm. The beam
       is then
       expanded with a bi-concave lens (f=-3cm) and then focused in
       the trap region by a bi-convex lens (f=15cm). The focal
       width is 2$\omega_{0}$=20$\mu$m.
       The photoionization probability per pulse~\cite{Monroe06}is
        \[
        P_{ion}=(I^{2}\sigma'/\hbar\omega)\tau
        \]
         where $\tau$ is the pulse length, 100fsec.
           So an estimate of the rate of photoionization averaged over the
          volume defined by the focal width of the beam and the length L of the trapping volume (which is smaller than the Rayleigh
          length) is
          calculated by averaging over the thermal trajectories of background
          atoms as they transverse the interaction volume
          to obtain
          \[
          R_{ion}=P_{ion}(n_{0}\omega_{0}/8T)(2\pi\omega_{0}L)
          \]
        where $n_{0}$ is the atomic density and T is the time between
         2 consecutive pulses. For a typical oven temperatures the pressure of the Sr vapor is $~10^{-9}$Torr
         which gives a estimate of $R_{ion}\thickapprox10sec^{-1}$.

        In separate "proof of principle" experiment the 431nm output beam was
        focused 5mm from the aperture of a small Sr oven in a vacuum chamber. Under the
        point where the atomic beam and the laser beam meet (2cm
        away) a electron multiplier(EM) was positioned
        to detect the ionised Sr atoms.
        The EM was  biased to $-2000V$ relative to a mesh 1cm above the photoionization region.
        The observed counting rate (fig.3)
        was consistent with theoretical predictions.
\begin{figure}[h]
\centering
\includegraphics[width=83mm]{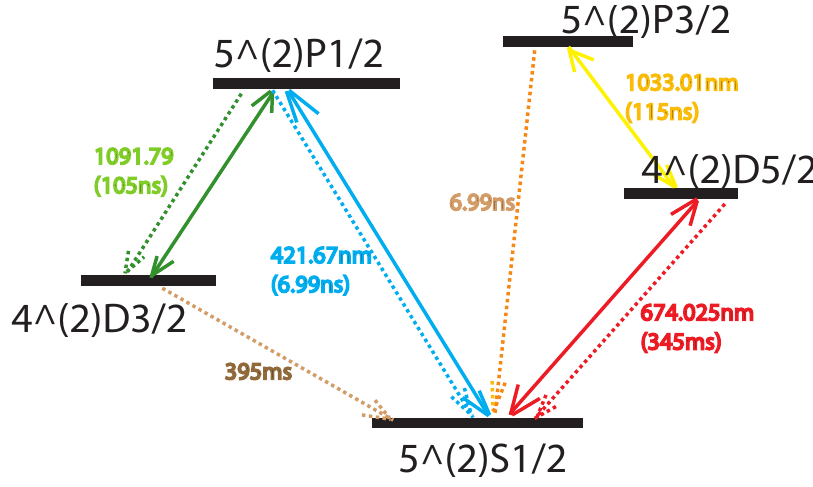}
\caption{Relevant levels structure.}
\end{figure}

\textbf{3. Laser systems for $Sr^{+}$}

 We constructed 4 laser systems required for driving the
 transitions in the "N-level" structure shown on fig.4.

\begin{figure}[h]
\centering
\includegraphics[width=123mm]{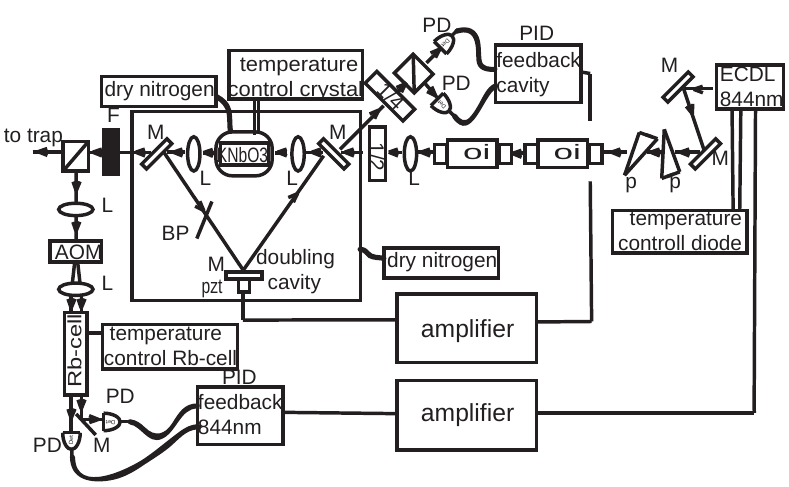}
\caption{422nm laser system for Doppler cooling driving the
$S_{1/2}\leftrightarrows P_{1/2}$ transition~\cite{Madej98}.}
\end{figure}

  The laser at 422nm (fig.5)~\cite{Madej98} excites the $S_{1/2}\leftrightarrows P_{1/2}$ cooling transition. It has a short
  life time of 7.87$sec$ and its fluorescence facilitates visual
  detection of even a single trapped ion.
  The heart of this laser system and also the one at 674nm is the
  ECDL (extended cavity diode laser).
\begin{figure}[h]
\centering
\includegraphics[width=53mm]{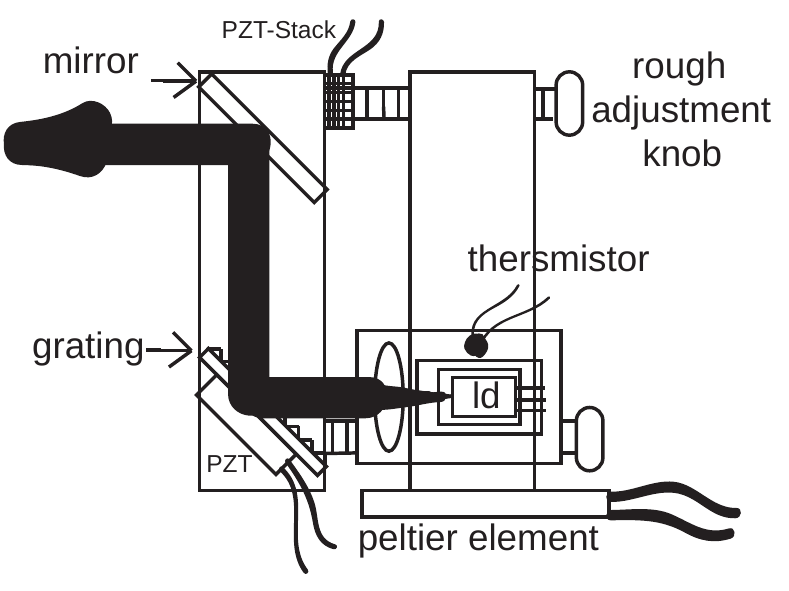}
\caption{Extended-cavity diode laser.}
\end{figure}

   We constructed custom designed ECDl's in a Littrow configuration
   (fig.6)~\cite{Scholten01}.
   An antireflection coated laser diode (100mW) provides a diverging elliptical beam centered at $\thicksim$850nm, which then passes through an aspheric
   collimating lens (0.55 NA,f=4.5 mm), mounted in a
   tube. The collimated light then falls on a diffraction grating (1800 lines/mm), which couples
   about 20\% of the light back to the laser diode
    and reflects ~75\% as a output. The whole structure lies on a modified mirror
    mount which is used for a rough adjustment of the diffraction
    grating. Precise positioning is provided by a PZT-stack mounted under
    one of the fine screws. The grating lies on a additional PZT piezoelectric transducer disk which is
    used to dynamically regulate the laser frequency based on an error
    signal (described below). A temperature sensor (10k$\Omega$) is
    used to detect the temperature close to the diode and a Peltier
    thermoelectric cooler is used to keep the temperature stable to
    1mK. The collimated beam next reflects from a mirror parallel to the grating
    so as to maintain a fixed output direction when the screw and the PZT stack are adjusted. The spectral width of
    such a ECDL is $\thicksim$1/2Mhz. The laser is encased in a
    1/2 inch thick aluminum box and lies on a heavy steel brick on top of a 4 sorbothane padding which enhances vibration isolation.

The collimated 844nm beam coming from the ECDL is circularized with
2 prisms and then passes through
     2 $\thicksim$30dB optical isolators. Next a bi-convex lens (f=150mm) is used to
     mode-match the beam into the doubling cavity and a $\lambda/2$
     plate is used to rotate the polarization.

     The triangular doubling cavity has 3 mirrors, a finesse of 60 and is 15cm on a side. The first mirror couples in the 844nm light with a 3$\%$
     transmission efficiency (S-polarized light, at 844nm and at $30^{\circ}$ angle of incidence), which is chosen to match the losses
     per turn in the cavity. The second mirror is the
     output coupler which for S-polarized light reflects $>$99.8$\%$, at
     $30^{\circ}$, for
     844nm and for P-polarized light transmits $>$80$\%$ at $30^{\circ}$ angle of incidence for
     422nm.
      The third mirror is simply a
     reflector $>$99.9$\%$ for 844nm, S-polarized light, at $30^{\circ}$ angle of
     incidence. A PZT-stack is positioned behind that mirror to provide feedback and keep the cavity
      locked to the 844nm laser (details below). A 1cm antireflection-coated $KNbO_{3}$ crystal is positioned halfway between, the input and output couplers,
      as well as
      2 bi-convex lenses (f=50mm, antireflection coated for 844nm). The lenses form a tight focus in the crystal,
     with Rayleigh length on the order of the crystal's length
      so as to efficiently generate the second
     harmonic ~\cite{chun88}. The crystal is noncritically, type I
     phasematched at $-17^{\circ}$ Celsius. Temperature is
     detected close to the crystal with AD590 sensor and is
     controlled, to 1mK precision with a Peltier element. The
     crystal, the Peltier element and copper heat sink (with a running chilled water) are in a acrylic chamber vented with dry
     nitrogen so as to, prevent freezing, preserve the
     antireflection coating and cavity finesse. The acrylic chamber is vented directly
     out of the cavity so as to avoid thermally driven changes in geometry. The
     doubling cavity also has a
     polarizing plate positioned at Brewsters angle which transmits only that light which is polarized parallel to the a-axis of the crystal. It is also
      a part of a modulation free  method for stabilizing the cavity~\cite{Hansch80}. This method is based on the measurement of the ellipticity
     of the light reflected from the input coupler. It uses a
     $\lambda/4$ plate, a polarizing beamsplitter with an axis at $45^{\circ}$ relative to the $\lambda/4$ plate and 2
     photodiodes. The  difference voltage generates an
     error signal which is then fed into the PZT-stack of the cavity. When the
     cavity contains a standing wave, which maximizes the pump (844nm)power in the
     crystal, the polarization of the light reflected from the input coupler is rotated
     from the polarization of the incoming 844nm light (but
     it is still linearly polarized). In this situation the 2 photodiodes detect
     equal signal. When any disturbance drifts the cavity away
     from resonance the reflected light is elliptically polarized
     and an error signal is created. The doubling system is enclosed
     in a
     sealed aluminum box and rests on sorbothane padding.

     Approximately 80$\mu W$ at
     422nm are produced from the enhancement cavity; 95$\%$ is directed into the trap and 5$\%$ is used to generate an absolute
     reference for the laser system~\cite{Madej98}. This stage
     begins with a
     blue pass filter (which eliminates the residue of 844$nm$
     light)and a f=150mm lens which focuses the light
     into an acousto-optic modulator. The AOM produces frequency
     downshifted and unshifted beams of similar power. The beams are collimated by a f=50mm lens
     so as to
     traverse a 7.5cm long Rb cell, stabilized at
     $130^{\circ}C$. The 2 beams are then detected by 2
     photodiodes.
      The Doppler-broadened absorption profile of the Rubidium vapor
     consists of 3 gaussian due to the natural mixture of $^{85}Rb$
     and $^{87}Rb$. The central gaussian is closest to the center of
     the $^{88}Sr^{+}$  $5s^{2}S_{1/2}\to 5p^{2}P_{1/2}$ transition
     (with a peak absorption $\thicksim$500Mhz lower than the targeted Sr transition). The width of
     the Rb absorption
     profile is $\thicksim$1.3Ghz. The down-shifted beam falls on the other side
     of the gaussian thus making the error signal (the difference of the signals of the 2 diodes) less sensitive to
     temperature drifts of the Rb cell and power drifts of the
     422$nm$ beam.

      To trigger the clock transition (fig.4)
      $S_{1/2}\leftrightarrows D_{5/2}$ a 674nm laser is needed (fig.7). This is a very narrow
      transition with lifetime 1/3sec which can be used to excite
      quantum jumps. It can also play the role of a "quantum bit", or serve as the
      analog of a spin up $|\uparrow\rangle$ and spin down
      $|\downarrow\rangle$ states of a spin 1/2 particle.
 \begin{figure}[h]
\centering
\includegraphics[width=123mm]{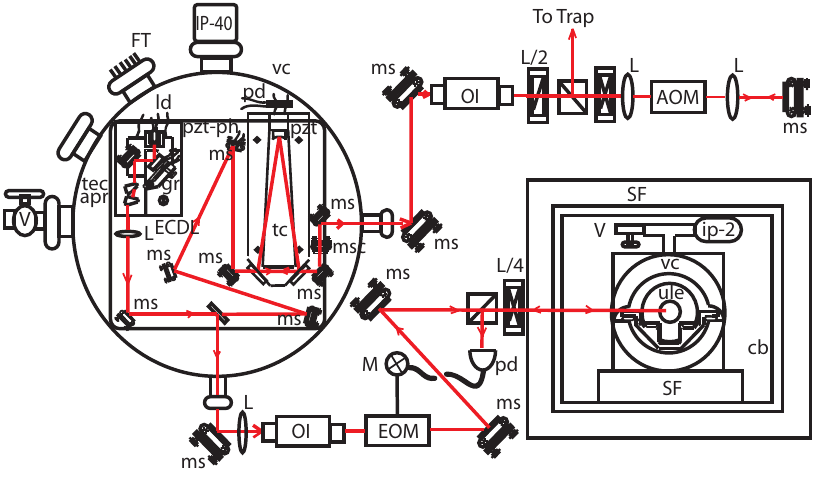}
\caption{Laser system for 674nm. $5^2S1/2\rightleftarrows4^2D5/2$
Strontium transition. The abbreviations used are: IP-ion pump,
FT-feedthrough, V-valve, VC-vacuum chamber, tc-triangular cavity,
ms-mechanical stages, pd-photodiode, pzt-piezoelectric crystal,
apr-anamorphic prisms, tec-Peltier element, kn-precision knob,
msc-optical coupler, OI-optical isolator, L/2-halfwave plate,
L/4-quarter wave plate, L-lens, AOM-aqcousto-optic modulator, EOM-
electrooptic-modulator, PD-avalanche photodiode, M-mixer,
SF-sterofoam isolation, cb-copper cage, ULE-Ulrta-low expansion
cavity
 }
\end{figure}
This is the most demanding laser system employed in the Strontium
ion trapping experiment.
       The spectral
       width of the laser is decreased to less than 10KHz in several stages.

       In the first stage the single antireflection coated diode
       at  670nm is placed in an ECDL (Extended Cavity Diode Laser)
       setup which is similar to the one used in the 422nm Doppler
       cooling laser system. Now however, the whole body
       of the ECDL including the mechanical stages are made from
       Super-Invar (thermal expansion $0.6\times
       10^{-6}(^{\circ}C)^{-1}$).
       The temperature of the ECDL is controlled to long-term
       stability of 1mK with a peltier element. After leaving the
       ECDL the beam with a spectral
       width of $~1Mhz$ is circularized with 2 prisms.
\begin{figure}[h]
\centering
\includegraphics[width=103mm]{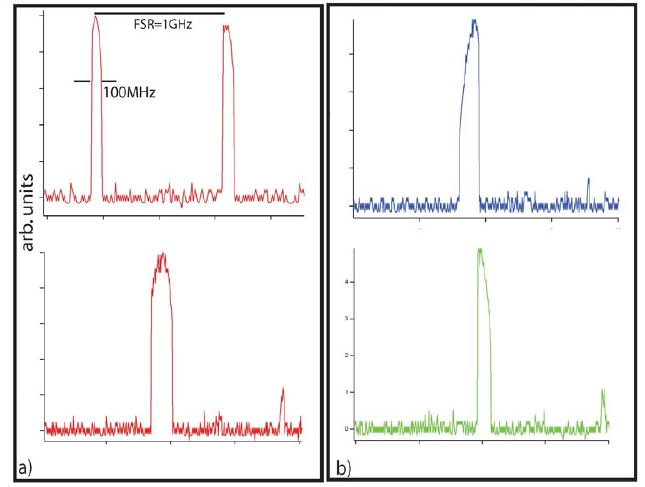}
\caption{Triangular cavity transmission a)cavity-pzt swept and
optical feedback path on resonance b)optical feedback out of
resonance, the peaks are asymetric.}
\end{figure}
       In the second stage the beam is mode-matched with a single
       lens to the $TEM_{00}$ mode of a triangular Fabri-Perot cavity
       with a FSR of 1Ghz and finesse of 200. Most of the designs
       implementing an optical feedback from a external cavity
       take the output before the cavity and need to adjust the
       cavity position so the first reflection from the input
       coupler doesn't couple back to the diode but the leakage beam
       does. This requires either a 2 mirror cavity in a almost
       confocal arrangement with slight misalignment or a triangular cavity which should be adjusted to couple back the leakage beam
       and to resonate at the same time ~\cite{Madej94}.
       In the design that we are using the optical feedback comes from a small $4\%$ reflective
       plate positioned after the output of the triangular cavity, which allows the cavity to be aligned and
       the optical feedback to be adjusted separately ~\cite{Chuang07}. Another advantage of this design is that the output is taken in transmission
        of the  triangular cavity so additional filtering is applied to the broad fluorescent
        background of the diode existing even after the optical feedback is applied. At the
       sharp angle of the triangular cavity there is a high
       reflector with a curvature R=75cm,
        chosen so the higher modes of the cavity
       are pushed away from the $TEM_{00}$ mode. Behind the
       reflector, which is mounted on a tubular
       pzt-stack a small photodiode detects the leakage from the
       reflector (fig.8). The path from the triangular cavity and the laser diode has to be
       adjusted so that the feedback beam and the outgoing one are in phase.
       For that purpose one of the reflectors guiding the beam between the ECDL and the cavity is glued on a
pzt-stack (called pzt-ph on fig.7)
        When the optical feedback is on the ECDL follows the cavity
       over a range of about 100Mhz (fig.8).
       Narrowing of the laser linewidth relative to the ECDL
       is then given by~\cite{Laurent89}:
       \[
\varDelta\nu=\frac{\varDelta\nu_{ecdl}}{\beta(\frac{L_p}{L_{ecdl}}\frac{F_{tc}}{F_{ecdl}})^2}
\]
where $L_p$ and $L_{ecdl}$ are the ECDL and cavity lengths and
$F_{tc}$ and $F_{ecdl}$ are the ECDL and cavity finesses. Linewidth
narrowing is given as the square of the ratio between the HWHM of
the triangular cavity transmission fringes (fig.8) with and without
optical feedback. We estimate that after the triangular cavity the
laser linewidth is $<10kHz$.

 The first and most important part of the laser system consisting of
 ECDL, triangular cavity and reflectors guiding the beam between the
 two is laid on a Super-Invar plate (8"x8"x1"), which is then
 positioned in a custom made UHV chamber (fig.7) held at
 $10^{-7}Torr$ only by a Starcell 40L/s ion pump. The super-invar
 plate lies on 4 viton o-rings and the whole vacuum chamber together with the ion pump sits
 on 2 stages of sorbothane padding in order to achieve
 sufficient vibration isolation.

 From the vacuum chamber the output in transmission of the
 triangular cavity which is spatially and spectrally filtered (even
 when the laser is narrowed from the optical feedback there is a
 broad fluorescent background from the laser diode) to send to the
 ion trap. To tune the beam across the Sr transition without
 deflecting the beam, we employ a double pass through a Acousto-optic
  modulator (Brimrose GPF-600-400).
\begin{figure}[h]
\centering
\includegraphics[width=83mm]{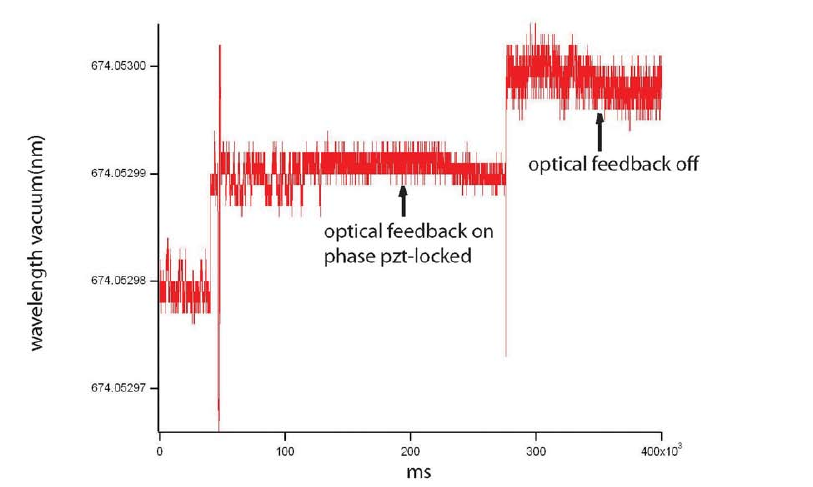}
\caption{Triangular cavity transmission a)cavity on resonance and
optical feedback path on resonance b)optical feedback off, spectral
linewidth determined just by the ECDL. }
\end{figure}

 A small fraction of the beam coming from the ECDL (about $60\mu
 W$) is deviated before reaching the triangular cavity and sent
 to a Ultra-low expansion cavity (ATfilms high grade Corning
 glass, $2\times 10^5$
 Finesse, 10cm length). The cavity is held in vacuum $10^{-7}Torr$
 in a temperature controlled custom made UHV chamber with a 2L/s ion
 pump acting on it. The vacuum chamber is then enclosed in a 1/4"
 thick copper box, temperature controlled with a nichrome heater
 wire to 2mK at $25^{\circ}C$(which is approximately the zero crossing temperature
 of the thermal expansion coefficient of the Corning ULE cavity).
 The copper box is further isolated from the room environment with a 1.5" thick
 styrofoam wall with small holes left only for the laser beams.
 The goal is to stabilize the ULE cavity temperature to $<10\mu K$.

 We intend to use Drever-Pound-Hall(DPH) method to lock the laser to the
 ULE cavity. To create a DPH error signal we modulate the laser phase
 with an Electro-optic modulator (New Focus) at 20Mhz. The error
 signal is detected in reflection (fig.7). The self-constructed Servo circuit
 used for feedback has a proportional gain of 0.5 and a integral gain of
 $10^6s^{-1}$. We intend to feed this signal to the injection
 current of the laser diode and a filtered slow component will be
 fed to the triangular cavity pzt.

 When the triangular cavity and the path between the triangular cavity and
 the ECDL are at resonance, so the transmission fringe looks like the
 bottom left part of fig.8,it
 takes an hour to observe a change bigger than $1\%$ even in the absence of a electronic lock .
  Such a lock has been implemented by dithering the "phase-pzt" to
  obtain transmission as a function of path length near the maximum.
  Derivatives of this signal determine the symmetry of the
  transmission peak. Feedback to the same "phase-pzt" then locks it
  to the path length that achieves symmetric transmission. The effect
  of this feedback control as recorded by the wavemeter is shown on
  fig.9.

 For more reliable long term stability and even narrower laser linewidth
 we will need the lock to the ULE cavity fringe although the performance at
 that stage is already suitable for sideband cooling experiments.

 To amplify the rate of sideband cooling, which will otherwise rely
 on the spontaneous decay of the $4^2D_{5/2}$ transition (1/3sec.)
 we built a laser at 1033nm which can open the $4^2D_{5/2}\leftrightarrows
 5^2P_{3/2}$ transition which has a short lifetime of 7ns down to the ground
 state. The 1033nm laser is a ECDL without an absolute reference
 but with good temperature control, vibration isolation and choice of
 materials (super-invar). It stays within 3Mhz for 1 hour, which
 is well within the transition bandwidth.

\begin{figure}[h]
\centering
\includegraphics[width=103mm]{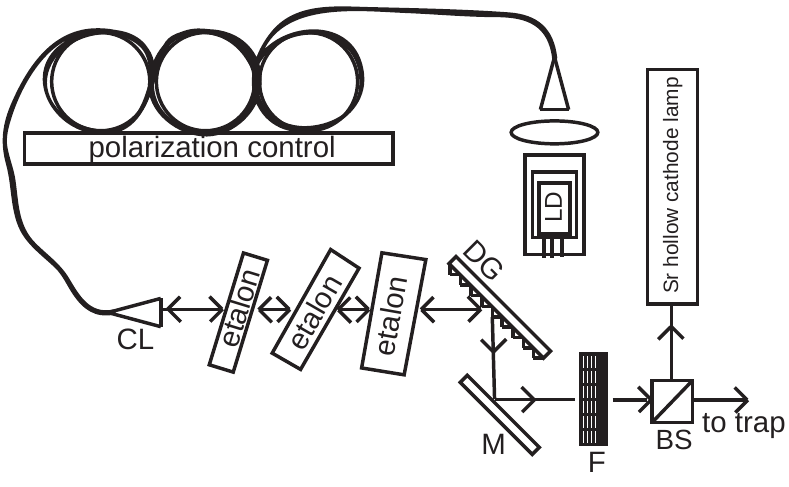}
\caption{1092nm laser system driving the $P_{1/2}\leftrightarrows
D_{3/2}$ transition.}
\end{figure}

         Unfortunately an ideal V-level structure in Sr doesn't exist and
         one in 13 decays of the $P_{1/2}$ level falls into the $D_{3/2}$
         state, so that laser cooling and
         fluorescence are interrupted. To optically pump the ions out of the
         $D_{3/2}$ level a laser at 1092nm is needed. We use a
         specially made $Nd^{3+}$-doped fiber (fig.10)~\cite{Madej89} laser. It is pumped
         by a 150mW laser diode at 830nm and it emits between 1060nm and 1100nm. The fiber core has 5$\%$
germania by eight containing $Nd^{3+}$-doped ions at circa
500-1000ppm and absorption is 3.5dB/m @ 830nm. The fiber (3m long)
is positioned between a input coupler which is antireflection coated
at 780-850nm range and is $>$99.9$\%$ reflective for 1060-1100nm.
One end of the fiber is contacted with the input coupler and on the
other end outgoing light is collimated to a 1mm beam which travels
$\thicksim$20cm before reaching the grating (1200lines/mm)
positioned at a littrow angle to roughly select the needed
wavelength. About $\thicksim$50$\%$ of the light returns back to the
cavity. Between the collimator and the grating there are 3 uncoated
etalons (1mm,5mm and 15mm thick) which are used to further narrow
the laser linewidth and for fine tuning. The full spectral width is
1Ghz and the tuning range is from 1070nm to 1100nm. The desired
wavelength, 1091.79nm can be simply dialed up on a wave meter, or
found by exciting an optogalvanic effect in a Sr hollow cathode
lamp. Active feedback is not needed because the spectral structure
of the laser consists of several modes spaced 40Mhz apart
(corresponding to the length of the cavity) and each one of them has
enough power (total power is 5mW) to saturate the transition. During
the time of the experiment the laser never drifts so dramatically to
leave the transition out of it's 1Ghz spectral width. Finally the
1092nm beam is overlapped with the 422nm and 674nm beams. The 431nm
beam for the photoionization is separate and perpendicular to the
other three. All the beams lie perpendicularly to the direction of
observation.

\textbf{4. Vacuum system}
\begin{figure}[h]
\centering
\includegraphics[width=123mm]{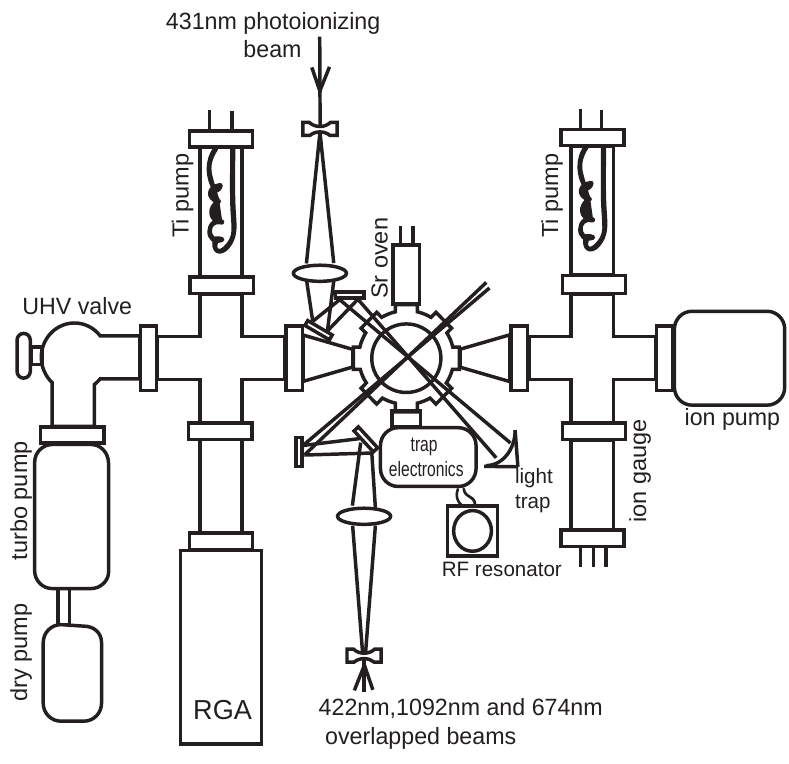}
\caption{Vacuum system and focusing optics.}
\end{figure}

 The vacuum system is shown in fig.11. The ion trap is positioned in
 the middle of a commercially available octagon with 2 (4.5inch)
 view ports used for detection of the ion fluorescence and 8
 (1.33inch) small ports used one for, electrical feedthrough, the Sr oven, the
 overlapped 422nm, 674nm and 1092nm beams, the 431nm
 photoionizing beam and pumping.

  The pressure, typically 8$\times10^{-11}$ Torr is achieved
  in several stages. First the range of ~2$\times10^{-8}$ is
  achieved
  by a turbo pump (20l/s), backed up by a diaphragm rough pump
  (capable of achieving 1-2Torr). The whole system is baked in place
  for 24 hours at ~150$^\circ C$(the only exception being the ion pump baked to 250$^\circ C$) after each time the chamber has been
  opened to atmosphere. Prior to the assembly, all steel components
  are baked at 450$^\circ C$ under vacuum for more than 1 weeks in
  order to minimize the hydrogen content, which is the factor that
  determines the lowest achievable pressure.

   In the second stage 2 titanium filaments are flashed for about 1min with
   50Amps of AC current to evaporate a titanium layer on the
   walls of the cylinders that surround the filaments. This reactive
   layer efficiently removes all the getterable gasses. The noble gases
   are pumped by a 20l/s ion Starcell pump.
    At this point a UHV valve is closed to separate the turbo pump
    from the rest of the experiment and only
    the titanium and ion pumps act on the chamber. The turbo and rough pumps are
    now switched off and pressure is monitored by an RGA
    and an ion gauge.

    \textbf{5. Imaging system}
\begin{figure}[h]
\centering
\includegraphics[width=83mm]{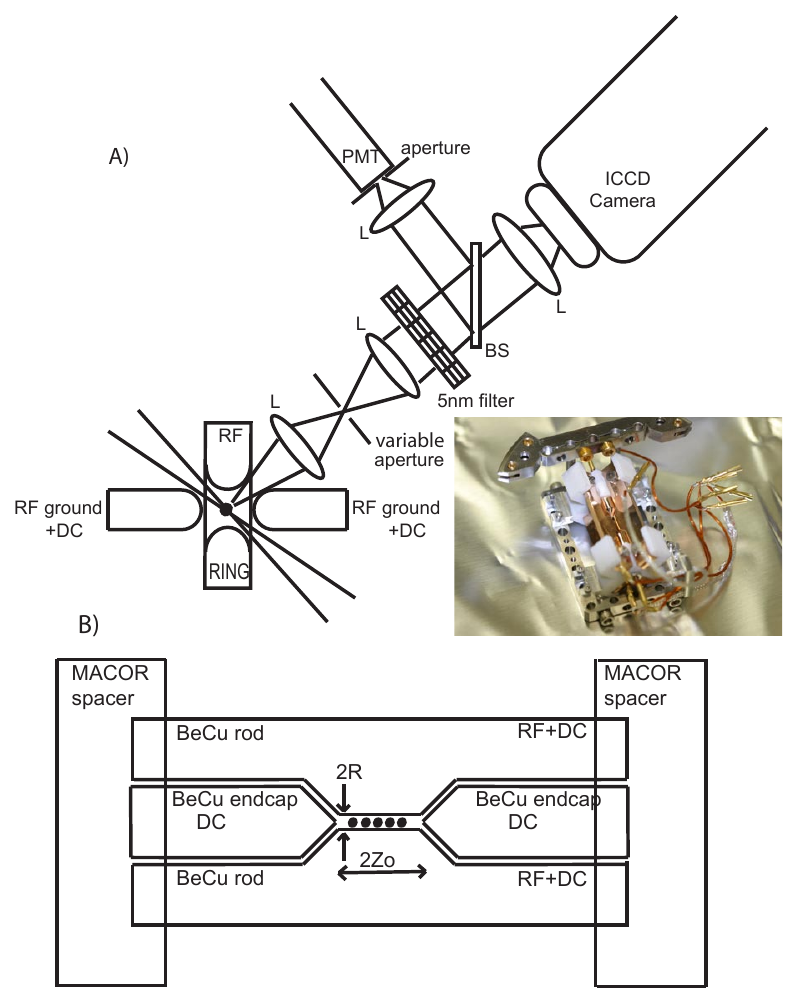}
\caption{a) Quadrupole Paul trap and imaging system b) Linear trap
(the same imaging system used).}
\end{figure}

     The imaging system (fig.12A) consists of a 28mm camera lens
     positioned above the top 4.5inch vacuum port. It creates a
     (1:1) image 5cm above the port. In the plane of the image
     an adjustable rectangular aperture (consisting of 4 independently translatable blades)is used to select the signal and suppress scattered light
     (which is especially strong from the electrodes).
     Next, the fluorescent light is collimated by a 28mm camera lens
and then refocused on the photocathode of a ICCD camera
      by a 200mm camera lenses. In between the 2 lenses there is a
      laser line filter centered at 422nm with width of 5nm. Also
      there is a 50/50 pellicle nonpolarised beamsplitter which
      deflects half of the collimated light onto a bi-convex lens (f=50mm),
      which focuses it onto a PMT tube. Such an arrangement allows for
      simultaneous observation of the ions with the camera and
      photon counting with the PMT.

      \textbf{6. Different Ion Traps}

      Fig.12a,b shows various ion traps that were built. We started with a standard Paul trap~\cite{Paul90,Wuerker59} with characteristic endcap
      distance $2z_{0}$=2mm
      (fig.8a). With this trap we used an iridium filament electron gun for ionization. The filament was
      biased at ~-40V relative to the endcaps and was positioned 7.5mm away from the trapping region. With 3Amps
running through the filament about 0.1$\mu$A of electron current
crossed the trap. The trap itself was made out of stainless steel
304. The only non-conducting component was made from Macor and
carefully screened. It was used to insulate the ring of the trap on
which the RF=300Vpp at 7Mhz was applied. On the other side of the
trap 1.5cm away, opposite to the e-gun a tantalum oven with several
1mm pieces of Sr was positioned. The oven has 1mm aperture and care
is taken to make sure that the atomic flux crosses the region of the
trap.

 We also built a linear trap(fig.12b)~\cite{Raizen92}. The distance between
 the endcaps is $2z_{0}$=6mm and the distance from the trap axis to the
 rod surface is R=0.75mm. The trap was machined out of
 Beryllium-Copper 172 alloy and the spacers were made from
 Macor as shown. With this trap we used
 photoionization (not the electron gun), so the exposed insulators were of less
 concern.
 Nevertheless the trap region doesn't directly "see" any insulating
 surface. The electrode surfaces were polished with 100nm grit diamond
 paste and then were chemically cleaned with sulphuric acid and hydrogen peroxide
 solution.
  The RF potential for both types of traps is provided by a quarter-wave
  helical resonator~\cite{macapin59} (Q=60) made out of copper with one secondary coil for
  the quadrupole Paul trap. For the linear trap 2 secondary coils
  provide the opportunity for separate DC offsets, which are used for
  compensation of the potential imperfections and to minimize micromotion of the ions. Separate DC potentials can also be applied to the
   endcaps and to the
  other side-rods that are at RF ground. Just outside the vacuum
  the DC electrodes are connected to ground with 0.1$\mu$F to
  reduce the inductively coupled RF noise.

\textbf{6. System performance}

\begin{figure}[h]
\centering
\includegraphics[width=83mm]{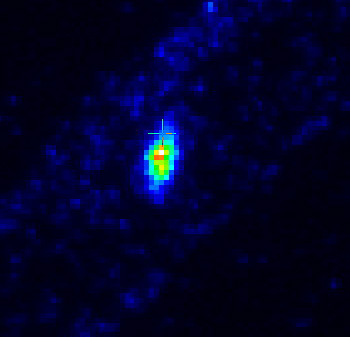}
\caption{Linear trap. A string of ions is formed in the trap. The
ions are visible one at a time due to the tight focus. Pixel size
~5x5$\mu m$. }
\end{figure}
 During an experimental run we first~\cite{berkeland2002}
 slowly lower the temperature of the $KNbO_{3}$ crystal (to
 prevent crack of the coating) until it reaches roughly
 $-17^{\circ}C$. During this time the temperature controllers
 for the ECDL's of the 422nm, 674nm laser and the Rb-cell are switched on.
 Thermal
 equilibrium is reached within 1/2 hour. At this point all the 4
 lasers are on with the light at 431nm being too weak to excite
 ionization, because the laser at 862nm is out of lock.

 To optimize the performance of the doubling cavity its PZT is
 pulsed at 10Hz and the transmission peaks at 844nm are
 observed on a removable photodiode. By adjusting the cavity
 components its finesse is maximized. At this point the
 diode is removed and the cavity is locked to the 844nm laser. Now
 the 2 beams at 422nm, 1092 nm are sent to the wavemeter and their
 wavelengths are adjusted by tilting the ECDL's cavities via the
 PZT-stacks.
  By sweeping the wavelength of the 422nm laser we
  can map out the absorbtion profile of the Rb vapor and lock the
  844nm ECDL to its proper location. The RF potential is switched on to 300Vpp and the DC potential on the endcaps is adjusted at 50V.
  The Sr oven has by now been preheated for about 2 hours with$\thicksim$1.5Amps
  DC current. The oven current is now increased to 2.5Amps and
  simultaneously the 431nm output is fired up as the 862nm laser is modelocked. Several amps running through a small coil
  under the vacuum chamber generate a magnetic field of several
  gauss
  which prevents the
  ions from being driven into a dark state~\cite{Berkeland002}.

After several seconds a small cloud is visible on the ICCD camera
and both the ionizing laser and the oven are switched off. Now we
gradually decrease the RF and the DC on the endcaps until most of
the ions escape the trap. The remaining ions form a stable linear
string (for the linear trap) that can live many hours (fig.13).

\textbf{7. Acknowledgments}

 It is a pleasure to thank Alan Madej and Pierre Dube(National research council of
 Canada) and Robert Scholten(University of Melbourne) for the numerous very helpful advices on the
 fabrication of the laser
 systems. Ralph Wuerker , Dana Berkeland (LANL) and Nan Yu (JPL) for experimental
 suggestions and recommendations on the whole experiment, Brian Naranjo (UCLA,
 Physics department) for providing his help and knowledge for the vacuum
 system construction, Makan Mohageg(UCLA,JPL) for advises on the PID circuits and Alex Bass (UCLA, Physics
 department) for providing deep theoretical ideas on fundamental
 physics questions which was rather encouraging. Research supported
 in part by Darpa and the ONR, and a gift from the Elwood and
 Stephanie Norris Foundation.


\end{document}